\documentclass[twocolumn,showpacs,preprintnumbers,amsmath,amssymb]{revtex4-1}
\usepackage{graphicx}
\usepackage{textcomp}

\begin{document}
\title{Observation of surface states with algebraic localization}
  \normalsize

\author{G. Corrielli$^{1,2}$, G. Della Valle$^{1,2}$, A. Crespi$^{2,1}$, R. Osellame$^{2,1}$, and S. Longhi$^{1,2\ast}$}
\address{$^1$ Dipartimento di Fisica - Politecnico di Milano, Piazza Leonardo da Vinci 32, 20133 Milan, Italy}
\address{$^2$ Istituto di Fotonica e Nanotecnologie - Consiglio Nazionale delle Ricerche, Piazza Leonardo da Vinci 32, 20133 Milan, Italy}

%
\bigskip
\begin{abstract}
We introduce and experimentally demonstrate a class of surface bound states with algebraic decay in a one-dimensional tight-binding lattice. Such states have an energy embedded in the spectrum of scattered states and are  structurally stable against perturbations of lattice parameters. 
 Experimental demonstration of surface states with algebraic localization  is presented in an array of evanescently-coupled optical waveguides with tailored coupling rates.  \noindent
\end{abstract}



\maketitle

Surface waves localized at an
interface between two different media play an important role in different areas of  physics  \cite{Davi}. 
A widespread belief is that surface waves are exponentially localized waves. Indeed, exponential localization is ubiquitous for evanescent waves.
Exponential localization is found for electrons at the surface of a periodic crystal, the so-called Tamm \cite{Tamm} and Shockley \cite{S} surface states with energy in a gap, in disordered lattices as a result of Anderson localization \cite{Anderson}, or  at a metal-dielectric interfaces in the form of plamonic waves. 
However, quantum mechanics does not exclude the existence of localized states with a {\it lower} than exponential localization. Sub-exponential localization, including a power-law decay of the wave function, can arise, for example, in lattice models with special kind of disorder \cite{d1,d2}. Surface states with algebraic localization were predicted 20 years ago in certain special potentials for the Schr\"{o}dinger equation on a semi-infinite line \cite{wow}. Such surface states have an energy embedded in the continuous spectrum of scattered states, i.e. they belong to the class of bound states in the continuum (BIC) originally discovered by von Neumann and Wigner in a seminal paper \cite{Wigner} and found in a wide range of quantum and classical systems,  including atomic and molecular  systems \cite{a1,a2,a3}, semiconductor and mesoscopic structures \cite{m0,m1,m1,m2,m3,m4,m6}, graphene \cite{gr}, quantum Hall insulators \cite{NC},  Hubbard models \cite{H1,H2}, and optical structures \cite{o1,o2,o3,o4,o5}. Experimental demonstrations of BIC states, either in the bulk \cite{o4} or at the surface \cite{o5}, have been recently reported in simple optical lattice systems exploiting destructive Fano interference. Such  BIC states  are compact, i.e. they confine all the energy in few sites  with no penetration into the lattice continuum, and are thus not suited to observe sub-exponential localization. Recently,  surface states with sub-exponential  localization have been theoretically introduced by Molina and coworkers in a special tight-binding lattice model \cite{Molina12}. Such states are BIC modes  which, as opposed to those earlier studied in Refs.\cite{o1,o4,o5}, are not compact and penetrate in the lattice with a sub-exponential (but higher than algebraic) localization. However, like in \cite{wow} a specially-tailored local potential is required, which is of difficult experimental implementation. The observation of surface states with sub-exponential localization  remains to date elusive.  \par
In this Letter we introduce and experimentally demonstrate surface states with power-law decay in a semi-infinite tight-binding lattice model, which do not require any local potential. Algebraic localization exploits the existence of a BIC mode, and it is not related to a special kind of disorder in the lattice \cite{d1,d2}.  Our scheme is experimentally demonstrated in an array of coupled optical waveguides with tailored hopping rates, manufactured by femtosecond laser writing in fused silica. Algebraic localization of the surface state is proven by spectral reconstruction of the BIC eigenmode from beam propagation measurements.\par We consider a semi-infinite tight-binding lattice with inhomogeneous hopping rates $\kappa_n$ and site energies $\epsilon_n$ ($n=1,2,3,...$) described by the tight-binding Hamiltonian 
\begin{figure}
\includegraphics[width=8cm]{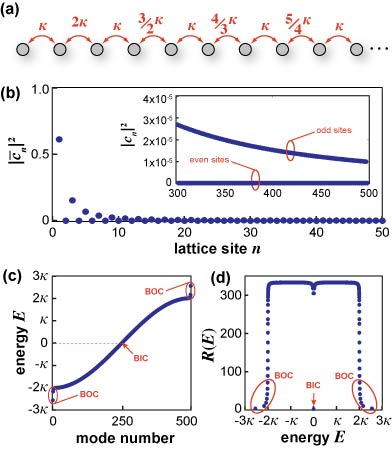}
\caption{(Color online) (a) Schematic of a semi-infinite lattice with tailored hopping rates that sustains $M=1$ BIC surface state with algebraic decay ($\beta=1$). (b) Behavior of $|\bar{c}_n|^2$ for the BIC state. (c) Numerically-computed energy spectrum of the lattice comprising $N_s=501$ sites, and (d) corresponding behavior of the participation ratio $R(E)$ of the eigenmodes. The BIC state has an energy $E=0$. BOC modes are located close to the band gap edges $ \pm 2 \kappa$. The outer BOC modes have an energy $ E=\pm E_0 \simeq  \pm 2.557 \kappa$.}
\end{figure}
\begin{equation}
\hat{H}=- \sum_{n=1}^{\infty}  \left\{ \kappa_n | n \rangle \langle n+1 |+ \kappa_{n-1} | n \rangle \langle n-1| \right\}+ \sum_{n=1}^{\infty} \epsilon_n |n \rangle \langle n |
\end{equation}
where $\kappa_n>0$ is the hopping rate between sites $|n \rangle$ and $|n+1\rangle$, and $\kappa_0=0$. We assume that, far from $n=1$, the lattice is homogeneous, i.e. $\epsilon_n \rightarrow 0$ and $\kappa_n \rightarrow \kappa$ as $n \rightarrow \infty$. The energy spectrum of $\hat{H}$ is obtained from the following eigenvalue equation for the occupation amplitudes $\bar{c}_n$ of various lattice sites
\begin{equation}
E \bar{c}_n=-\kappa_n \bar{c}_{n+1}-\kappa_{n-1} \bar{c}_{n-1}+\epsilon_n \bar{c}_n
\end{equation}
($n=1,2,3,...$). The linear spectrum of scattered states, i.e. the continuous spectrum of $\hat{H}$, is provided by the tight-binding lattice band $-2 \kappa <E < 2 \kappa$. Bound  states can arise owing to the inhomogeneity of the hopping rates $\kappa_n$ and/or of the local potential $\epsilon_n$. A method to create a single BIC surface state in a lattice with $\kappa_n=\kappa$ and with a specially tailored local potential $\epsilon_n$ was proposed in Ref.\cite{Molina12}. Here we suggest a different and experimentally more accessible method to synthesize a discrete lattice that sustains an arbitrary number $M \geq 1$ of surface BIC with algebraic localization that does not require any local potential, i.e. $\epsilon_n=0$. Our idea is to introduce a modulation of the hopping rates $\kappa_n$ between adjacent sites, which can be simply realized in a semi-infinite tight-binding lattice with inhomogeneous spacing of adjacent lattice sites. Some general properties of the Hamiltonian $\hat{H}$ in the $\epsilon_n=0$ case and for inhomogeneous hopping rates are discussed in the Supplemental Material \cite{supp}.
 To realize a BIC state with algebraic localization, let us modulate the lattice hopping rates $\kappa_n$ as follows
 \begin{equation}
\kappa_n= \left\{ 
\begin{array}{ll}
\kappa &  n \neq l N \\
\left( \frac{l+1}{l} \right)^{\beta}\kappa & n=lN \; \; (l=1,2,3,...)
\end{array}
\right.
\end{equation}
where $N=M+1$ and $\beta$ is an arbitrary real number that defines the power-law decay exponent ($\beta>1/2$ for normalizable states). 
Indeed, it can be readily shown that Eqs.(2) admit of the following $M$ surface states 
\begin{equation}
\bar{c}_n^{(\sigma)}=A_n \sin (n q_{\sigma}) 
\end{equation}
with energies $E_\sigma=-2 \kappa \cos q_{\sigma} $ buried in the band of scattered states. In the previous equation, $\sigma=1,2,...,M$,  $q_{\sigma}= \pi \sigma /N$,  $A_n= \mathcal{N}l^{-\beta}$ for $ (l-1)N < n \leq lN$ ($l=1,2,3,...$), and $\mathcal{N}$ is a normalization constant.  As an example, in Fig.1(a) we show the very simple discrete lattice that sustains one surface BIC with the algebraic decay law $\bar{c}_n \sim 2/(n+1)$, i.e. corresponding to $N=2$ and $\beta=1$, with energy $E=0$ at the center of the tight binding lattice band. The distribution of the surface BIC is depicted in Fig.1(b). In addition to surface BIC, the lattices defined by the sequence (3) sustain  additional surface states in the gap, i.e. bound states outside the continuum (BOC).  As an example, in  Fig.1(c) we show the numerically-computed energy spectrum of Eq.(2) in a lattice comprising $N_s=501$ sites for $N=2$ and $\beta=1$, i.e. for the lattice shown in Fig.1(a). The degree of localization of the
eigenstate $\bar{c}_n(E)$ with energy $E$ is measured by the participation ratio $R(E)$, given by $R(E)=(\sum_n| \bar{c}_n|^2)^2/(\sum_n |\bar{c}_n|^4)$ \cite{Molina12}. For localized modes, $R \sim 1$ while
for extended states $R \sim N_{s}$. The distribution of $R(E)$ for the $N_s=501$ eigenmodes of the lattice of Fig.1(a) is shown in Fig.1(d). The figure clearly shows the existence of one BIC surface state at $E=0$, together with a number of  BOC surface states (26 for the truncated lattice with $N_s=501$ sites) with exponential decay tails and with energies outside the lattice band. The two outer BOC states have an energy $ E \sim \pm E_0 \simeq  \pm 2.56  \kappa$, whereas the energies of the other BOC modes condensate toward the band gap edges $E= \pm 2 \kappa$. The surface BIC turns out to be structurally robust against perturbations of lattice parameters, as discussed in Ref. \cite{supp}. \par
\begin{figure*}
\includegraphics[width=17cm]{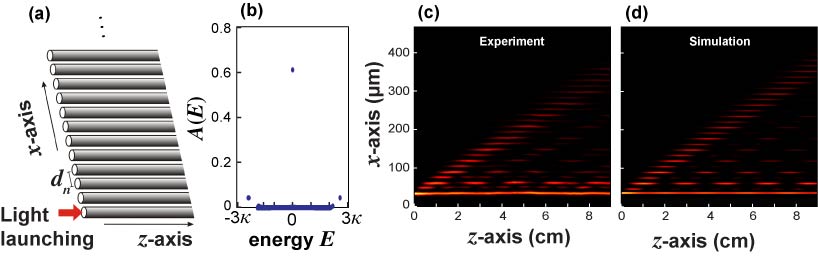}
\caption{(Color online)  (a) Schematic of the waveguide lattice; the red arrow indicates the waveguide where light is launched. (b) Mode excitation amplitude $A(E)$ for initial lattice excitation at the boundary site, i.e. $c_n(0)=\delta_{n,1}$ (c) and (d) Experimental and theoretical maps of the light intensity evolution in the lattice waveguides.}
\end{figure*}
 To experimentally demonstrate surface BIC modes, we implemented the semi-infinite lattice of Fig.1(a)  in an array of 40 evanescently-coupled optical waveguides manufactured by femtosecond laser waveguide writing on a fused silica substrate (see, for instance, \cite{refsper1,Rabi}). The second harmonic of an Yb-based femtosecond
laser (FemtoREGEN, HighQLaser GmbH), delivering
400 fs pulses, is used for the writing process. An
optimal processing window was found at 20 kHz repetition
rate, 300 nJ pulse energy and 10 mm/s translation speed.
The laser beam is focused at 170 $ \mu$m  below the glass surface by a 0.45 NA, 20 $ \times$ objective.
 The spacing $d_n$ between waveguide $|n \rangle$ and $|n+1 \rangle$ is engineered 
in order to implement the desired coupling coefficients, namely $\kappa_n / \kappa= 1,2,1,3/2,1,4/3,1,5/4,...$. For our waveguide writing parameters,  the coupling constant $\kappa_n$ turns out to be well fitted by the exponential curve $\kappa_n=\kappa \exp[-\gamma (d_n-a)]$, where $\kappa=1.27$ cm$^{-1}$ is the coupling constant for a waveguide spacing $a=15 \; \mu$m and $\gamma=0.20 \; \mu$m$^{-1}$. The values of $\kappa_n$ of the lattice of Fig.1(a) are obtained with spacings in the range $d_n=11.5 \div 15 \; \mu$m. The array was probed at $\lambda=633$ nm from light emitted by a He-Ne laser. Note that in our optical setting the spatial light propagation along the axial distance $z$ of the array reproduces the temporal evolution of the occupation amplitudes $c_n(t)$ in the lattice model described by the Hamiltonian (1), with $t=z$. To prove the existence of the surface BIC mode with algebraic localization, we measured the propagation of a light beam in the arrayed structure under suitable excitation at the input plane and used a spectral method to reconstruct  the eigenenergy and profile of the BIC mode \cite{spec1}.  The method basically requires to measure 
the correlation function $C(t)$ of the evolving optical wave packet $|\psi(n,t) \rangle=\sum_n c_n(t) | n \rangle$, i.e. $C(t)=\langle \psi(0) | \psi(t) \rangle$, and the evolution of $c_n(t)$ in the various lattice sites. Fourier analysis of the correlation function enables  to localize the position of the discrete eigenvalues as resonance peaks, whereas a Fourier analysis of $|\psi(n,t) \rangle$ generates the eigenfunction profiles \cite{spec1}. Technical details of the spectral method are given in Ref.\cite{supp}. To correctly reconstruct the eigenvector corresponding to the BIC state of the Hamiltonian (1), two conditions should be met: (1) the initial wave packet should have a non-negligible overlap with the BIC mode; and (2) the wave packet evolution should be monitored for a time $T$ much longer than $\sim 1 / \kappa$.  The latter condition arises because the BIC state is embedded into the spectrum of scattered states, whose contribution into the reconstructed state should be avoided.  Once the two conditions (1) and (2) are met, the energy position $E=E_1$ of the BIC state is found as a resonance peak of the Fourier transform $\hat{C}(E)$  of $C(t)$, i.e.
\begin{equation}
\hat{C}(E)=\int_{- \infty}^{\infty} dt g(t) \langle \psi(0) | \psi(t) \rangle \exp(iEt)
\end{equation}
whereas the corresponding eigenvector, apart from a normalization factor, is reconstructed via the relation 
\begin{equation}
\bar{c}_n(E_1) \simeq \int_{-\infty}^{\infty}dt g(t) c_n(t) \exp(iE_1 t).
\end{equation}
\begin{figure}
\includegraphics[width=8.5cm]{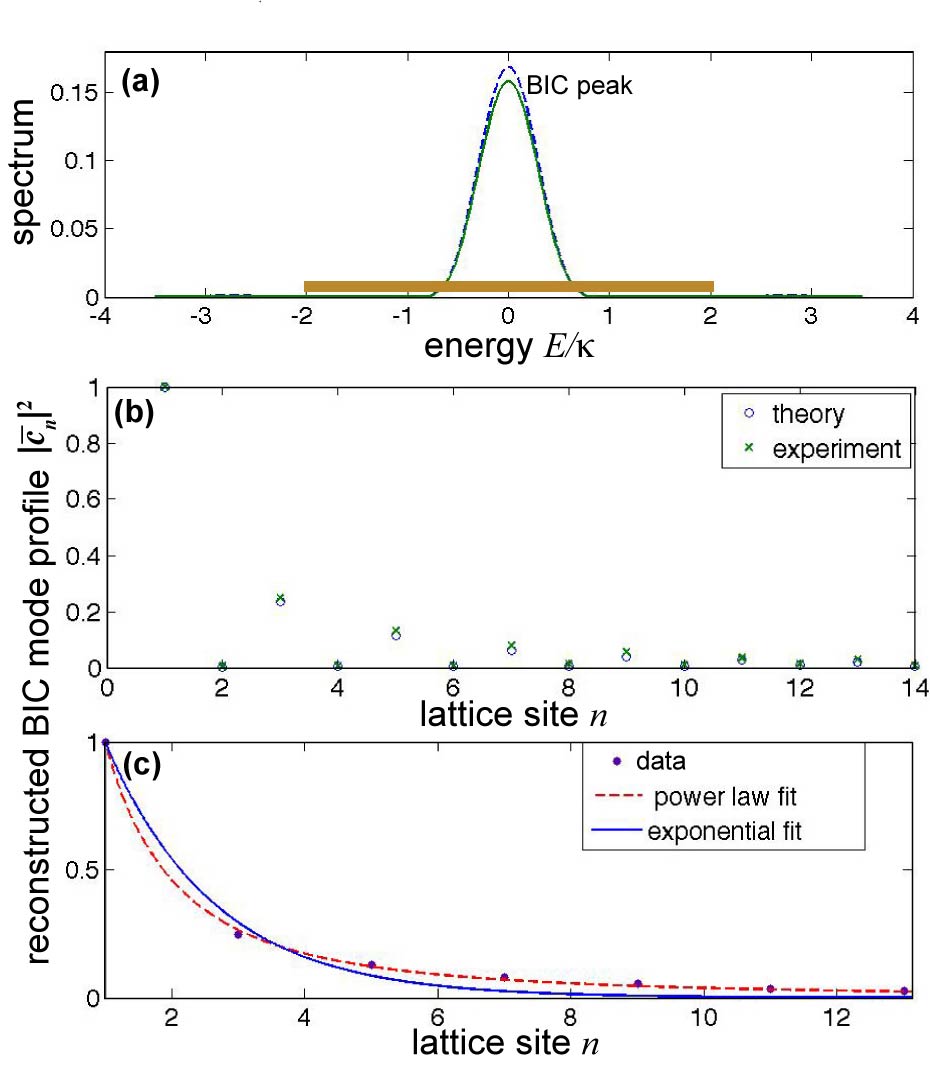}
\caption{(Color online) (a) Spectrum of the autocorrelation function $|\hat{C}(E)|^2$ versus normalized energy $E/\kappa$ as obtained from the experimental data (solid curve) and from the theoretical analysis (dashed curve). The solid horizontal line marks  the continuous spectrum of the lattice band. The peak at $E=0$ corresponds to the BIC mode. (b) Reconstructed BIC mode profile $|\bar{c}_n(E=0)|^2$ (crosses refer to the experimental data, open circles to numerical simulations). (c) Best exponential (solid line) and algebraic (dashed line) fitting curves of $|\bar{c}_n(E=0)|^2$  at odd lattice sites. The algebraic fitting curve is $|\bar{c}_n|^2=1/[(n+1)/2]^{2 \alpha}$ with best fit $\alpha=0.96$.}
\end{figure}
In Eqs.(5) and (6), $g(t)$ is a window function of length $ \sim T$, which can be chosen to be a Gaussian or a square-wave function \cite{spec1,supp}.  Its Fourier transform is the spectral filtering function, whose spectral width $ \sim 1 /T$ sets the minimum separation in
energy levels that can be resolved \cite{spec1}. 
In our experiment, to meet the above mentioned conditions we propagated light along the array for a distance of $9$ cm, corresponding to a time $T \simeq 11.4 / \kappa$, and excited the system at the left boundary waveguide $n=1$, see red arrow in Fig.2(a), corresponding to the initial wave packet $|\psi(n,0) \rangle= |1 \rangle$. 
For such an initial condition, the various lattice eigenmodes are excited with a weight $A(E)$ which is depicted in Fig.2(b).  The figure clearly shows that the BIC state with energy $E=0$ is the most excited eigenstate, and the condition (1) above is met. Moreover, the propagated time $T \simeq 11.4 / \kappa$ is long enough to provide a satisfactory resolution of the BIC eigenvalue and a negligible contribution of the scattered states in the reconstruction of the BIC mode profile \cite{supp}, according to the condition (2).
As discussed in Ref.\cite{supp}, for the lattice Hamiltonian (1) with $\epsilon_n=0$ and for the chosen initial condition, $c_n(t)$ turn out to be either real-valued or purely imaginary-valued. Such a property greatly simplifies the experiment because a measurement of the light intensity distributions $|c_n(t)|^2$ is sufficient to retrieve the behavior of $c_n(t)$, and hence the computation of the correlation function and reconstruction of the eigenvector according to Eqs.(5) and (6). In our experiment, the evolution of $|c_n(t)|^2$  was measured by top-view imaging of the fluorescence signal emitted by the waveguides where red light is propagating \cite{Rabi,Szameit,NCo}; further details are given in Ref.\cite{supp}.  In Fig.2(c) we show the measured map of light intensity evolution along the 9-cm-long waveguide array. For comparison,  the corresponding theoretical map is also shown in Fig.2(d). Note the good agreement between the two maps. From the measured intensity map, we extracted the evolution of $c_n(t)$ in the various guides \cite{supp}, and computed the spectrum of the correlation function using a Gaussian filter $g(t)=g_G(t)=(1/T) \exp[-(t-T/2)^2/w^2]$, truncated at $t<0$ and $t>T$, with $w=2T/5$. As compared to the square-wave filter, the Gaussian one avoids the appearance of oscillatory tails in the resonance peak, which might be erroneously attributed to the occurrence of other bund states. The computed spectrum $|\hat{C}(E)|^2$ of the autocorrelation function is  shown in Fig.3(a), clearly  indicating a resonance peak at $E=E_1=0$, i.e. the existence of a bound state with energy $E_1$ embedded into the spectrum of scattered states. The outer BOC modes are very weakly excited by the initial wave packet and thus they are not visible in the  spectrum of Fig.3(a). In Fig.3(b) we show the behavior of the reconstructed BIC eigenvector, obtained from Eq.(6) with $E=0$, for lattice sites up to $n=14$. In this case a square filter $g(t)$ has been used, which provides a slightly better estimate than the Gaussian filter. As expected, the values of $|\bar{c}_n|^2$ of even-number waveguides are  very small, whereas $|\bar{c}_n|^2$ at odd-number waveguides shows a slow decay. The results shown in Fig.3(a) and (b) are compared with the theoretical predictions based on Eqs.(5) and (6), where the amplitudes $c_n(t)$ are obtained by numerical integration of the coupled-mode equations rather than from the measured intensity map. To prove the algebraic (rather than exponential) localization of the BIC state, the reconstructed mode  amplitudes $|\bar{c}_n|^2$ are fitted, at odd lattice sites, by  
either an inverse power-law  curve $|\bar{c}_n|^2=1/[(n+1)/2]^{ 2 \alpha}$ or by an exponential curve $|\bar{c}_n|^2=\exp[-\alpha (n-1) ]$ with a single fitting parameter $\alpha$. The optimal fitting curves, obtained by minimizing the root mean square deviation (RMSD), are shown in Fig.3(c). The algebraic curve turns out to provide a much better RMSD than the exponential one (RMSD  0.01006 vs 0.04316), with an optimum fitting parameter $\alpha=0.96$, which deviates from the expected value $\alpha=1$ by $\sim 4 \%$.

In conclusion, we have introduced and experimentally demonstrated a new class of surface bound states with algebraic localization. While exponential localization is ubiquitous for evanescent waves, algebraic localization is found when the lattice sustains bound states with an energy buried in the spectrum of scattered states. Here a design procedure of  surface states with algebraic decay has been proposed for a tight-binding lattice model with inhomogeneous hopping rates and demonstrated using optical waveguide arrays.  Our results provide the first observation of surface states with algebraic localization in a controllable physical system and are expected to be of relevance to other fields, including ultracold atoms in optical lattices, electronic transport in quantum dot chains and mesoscopic structures, as well as other photonic systems.\\ 
\\
This work was supported by the European Union through the project FP7-ICT-2011-9-600838 ({\it QWAD - Quantum Waveguides Application and Development}).\\
\\
$^*$ Corresponding author: stefano.longhi@polimi.it

\newcommand{\rey}{\ensuremath{\mathrm{Re}}}
\newcommand{\Fig}[1]{Fig.~\ref{#1}}
\newcommand{\fig}[1]{fig.~\ref{#1}}
\newcommand{\Eq}[1]{Eq.~\ref{#1}}
\newcommand{\eq}[1]{eq.~\ref{#1}}
\newcommand{\bs}{\begin{split}}
\newcommand{\es}{\end{split}}
\newcommand{\e}{{\rm{e}}}

\newcommand{\rA}{{\rm{A}}}
\newcommand{\rB}{{\rm{B}}}
\newcommand{\rAB}{{\rm{A   B}}}
\newcommand{\rS}{{\rm{S}}}
\newcommand{\rP}{{\rm{P}}}

\newcommand{\rF}{{\rm{F}}}
\newcommand{\rI}{{\rm{I}}}
\newcommand{\rl}{{\rm{L}}}

\newcommand{\rmm}{{\rm{m}}}
\newcommand{\ri}{{\rm{i}}}
\newcommand{\rj}{{\rm{j}}}
\newcommand{\rk}{{\rm{k}}}
\newcommand{\rJ}{{\rm{J}}}
\newcommand{\rK}{{\rm{K}}}
\newcommand{\rL}{{\rm{L}}}
\newcommand{\rc}{{\rm{c}}}
\newcommand{\br} {{\bf{r}}}


\newpage 

\begin{center}
\section*{ Supplemental Material}
\end{center}
\renewcommand{\thesubsection}{S}
\renewcommand{\theequation}{S-\arabic{equation}}
\setcounter{equation}{0}

{\it {\bf S.1. Properties of the lattice Hamiltonian.}}
Let us consider the tight-binding lattice Hamiltonian, given by Eq.(1) in the main text, for $\epsilon_n=0$
\begin{equation}
\hat{H}=- \sum_{n=1}^{\infty}  \left( \kappa_n | n \rangle \langle n+1 |+ \kappa_{n-1} | n \rangle \langle n-1| \right)
\end{equation}
with $\kappa_0=0$, $\kappa_n>0$ for $n \geq 1$, and $\kappa_n \rightarrow \kappa$ as $ n \rightarrow \infty$. The state vector $|\psi(t) \rangle = \sum_{n=1}^{\infty} c_n(t) | n \rangle$ of the system evolves according to the Schr\"{o}dinger equation (with $\hbar=1$)
\begin{equation}
i \partial_t | \psi(t) \rangle = \hat{H} | \psi(t) \rangle,
\end{equation}
i.e. the following coupled-mode equations hold for the amplitudes $c_n(t)$ in the Wannier basis representation $\{ |n \rangle \}$
\begin{equation}
i \frac{dc_n}{dt} =- \kappa_n c_{n+1}-\kappa_{n-1} c_{n-1}.
\end{equation}
The following general properties hold.\\

Prop. I. The energy spectrum $E$ of $\hat{H}$ is symmetric around $E=0$.\\ 
In fact, if $| \phi_1 \rangle= \sum_n \bar{c}_n |n \rangle$ is an eigenstate of $\hat{H}$ with energy $E_1$, i.e. $\hat{H} | \phi_1 \rangle=E_1 | \phi_1 \rangle$, then it can be readily shown that $| \phi_2 \rangle= \sum_n (-1)^n \bar{c}_n |n \rangle$ is an eigenvector of $\hat{H}$ with energy $E_2=-E_1$.\\ 

Prop. II. The continuous spectrum of $\hat{H}$ is dense in the interval $(-2 \kappa, 2 \kappa)$.\\ This property follows from the fact that the semi-infinite lattice is asymptotically homogeneous. The asymptotic behavior of the scattered state $ | \phi(E) \rangle =\sum_n \bar{c}_n | n \rangle$ with energy $E$ as $n \rightarrow \infty$ is obtained from the asymptotic form of the eigenvalue equation $\hat{H} | \phi(E) \rangle=E | \phi(E) \rangle$ at $n \rightarrow \infty$, i.e. $-\kappa(\bar{c}_{n+1}+\bar{c}_{n-1})=E \bar{c}_n$. The most general solution of the asymptotic equation is given by 
\begin{equation}
\bar{c}_n \sim  A(q)[ \exp(iqn)+r (q)\exp (-iqn)], 
\end{equation}
where $r(q)$ is the reflection coefficient ($|r(q)|=1$ for flux conservation), $ 0 \leq q \leq  \pi$ is the wave number of the scattered state with energy $E=E(q)$ defined by the dispersion relation  
\begin{equation}
E(q)=-2 \kappa \cos q,
\end{equation}
and $A(q)$ is a normalization factor (for scattered states). The asymptotic form of scattered states given by Eq.(S-4) is valid provided that $A(q)$ does not vanish. This exceptional circumstance might occur at some energies $E$ whenever, starting from the asymptotic form (S-4) and propagating the amplitudes $\bar{c}_n$ backward with respect to the index $n$ using e.g. a transfer matrix method, a secular growth of $|\bar{c}_n|$ is found, which implies $A(q) \rightarrow 0$ for boundedness.  At such special energies the scattered states do not have the form given by Eq.(S-4) and might become normalizable, i.e. BIC modes can be found. This case can occur rather generally at $E=0$, as shown in the next proposition. For the inhomogeneous lattices considered in our work [see Eq.(3) given in text],   missing of scattered  states of the asymptotic form given by Eq.(S-4) and the appearance of BIC states occurs for a finite set of energies.\\

Prop. III. The even-number occupation amplitudes of the eigenstate of $\hat{H}$ with energy $E=0$ vanish, and the eigenstate may correspond to a normalizable state (i.e. a  BIC state)  with sub-exponential localization or to a scattered (non-normalizable) state of $\hat{H}$, depending on the asymptotic behavior of the sequence $\{ \kappa_n \}$.\\ In fact, the eigenvalue equation $\hat{H} | \phi \rangle=E | \phi \rangle$ for $E=0$ yields $\kappa_n \bar{c}_{n+1}+\kappa_{n-1} \bar{c}_{n-1}=0$, which is satisfied by taking (apart from a normalization factor) $\bar{c}_n=0$ for $n $ even, $\bar{c}_1=1$, and $\bar{c}_{n+1}=-(\kappa_{n-1}/ \kappa_{n}) \bar{c}_{n-1}$ ($n=2,4,6,...$).  Note that $|\bar{c}_{n+1} / \bar{c}_{n-1}|^2=(\kappa_{n-1}/\kappa_n)^2 \rightarrow 1$ as $n \rightarrow \infty$, so that  from the ratio test of convergence the norm of the eigenvector $\langle \phi | \phi \rangle=\sum_{n=1,3,5,...}|\bar{c}_{n}|^2$ may or may not converge. In the former case the eigenstate corresponds to a BIC state and,  since $|\bar{c}_{n+1} / \bar{c}_{n-1}|^2 \rightarrow 1$ as $n \rightarrow \infty$, the localization is sub-exponential. In the latter case the eigenstate is not normalizable and belongs to the continuous spectrum of $\hat{H}$, despite its form is not given by Eq.(S-4). For example, assuming for $\kappa_n$ the form given by Eq.(3) in the main text, the eigenstate is normalizable (i.e. it corresponds to a BIC mode) for $\beta>1/2$, whereas it is a scattered state for $\beta<1/2$.\\

Prop. IV. For the initial lattice excitation $ | \psi(0) \rangle =\sum_{n} c_n(0) |n \rangle$, with $c_n(0)$ real numbers for $n$ odd (even) and $c_n(0)$ purely imaginary numbers for $n$ even (odd), then the solutions $c_n(t)$ to the coupled-mode equations (S-3) remain alternately real and imaginary  numbers, i.e. $c_n(t)$ are real numbers for $n$ odd (even) and $c_n(t)$ purely imaginary numbers for $n$ even (odd) at any successive time $t$.\\ 
Such a property follows straightforwardly from an inspection of Eqs.(S-3). \\

{\it {\bf S.2. Structural stability of the BIC mode}}. An important issue for the experimental observation of a BIC mode is its structural stability [1], that is, whether the mode is stable against lattice perturbations of the hopping rate and site energy distributions caused, for example, by imperfections in waveguide spacing and refraction index distribution of waveguides. We checked the structural stability of the BIC mode with energy $E=0$ by computing the energies and corresponding distribution of eigenmodes of a lattice which differs from the target one, shown in Fig.1(a) given in the text, by adding perturbations $\delta \kappa_n$ and $\delta \epsilon_n$ to the hopping rates and site energies, respectively. The results obtained assuming for  $\delta \kappa_n$ and $\delta \epsilon_n$ random numbers uniformly distributed in the interval $(-0.05 \kappa,0.05 \kappa)$, i.e. for a moderate disorder of $\sim 10 \%$, are shown in Fig.4. The curve of the participation ratio $R(E)$, shown in Fig.4(d), clearly indicates the persistence of one localized mode with an energy $E \simeq -0.0137 \kappa$, which is slightly shifted from $E=0$. The distribution of $|\bar{c}_n|^2$ of the BIC state is shown in Fig.4(e). The other eigenmodes with energies inside the lattice band tend  to show a certain degree of localization due to the disorder (Anderson localization), as discussed in Ref.[1].\\

\begin{figure}
\includegraphics[width=8cm]{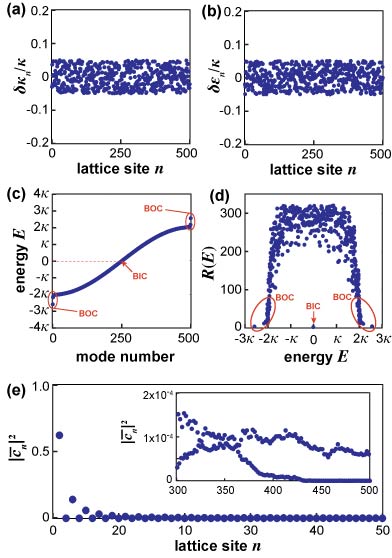}
\caption{(Color online) BIC surface mode in a lattice with disorder. (a) and (b) show a disorder realization of hopping rates $\delta \kappa_n$ and site energies $\delta \epsilon_n$. (c) Energy spectrum of the disordered lattice ($N_s=501$ sites). (d) Participation ratio $R(E)$ of lattice eigenmodes. (e) Behavior of $|\bar{c}_n(E)|^2$ for the BIC localized eigenmode with  energy $E=-0.0137 \kappa$.}
\end{figure}

{\it {\bf S.3 Spectral Method}.} Spectral methods  enable rather generally to reconstruct the discrete spectrum and the profile of the corresponding eigenvectors of an Hamiltonian $\hat{H}$ from the Fourier  analysis of the temporal evolution of a wave packet [2,3]. Such methods are useful tools from a computational viewpoint [2,3], for example for an accurate estimate of eigenvalues and eigenvectors of the Schr\"{o}dinger equation, as well as for the analysis of experimental data in systems where the evolution of a wave packet can be accessed. 
In our optical experiment, the wave packet evolution  is visualized by simply monitoring the propagation of a light beam along the waveguides of the array, as discussed in the main text.\\ 
The method that we used to analyze our experimental data is a simple spectral filtering method, which is discussed in Ref.[2]. The method is limited by the uncertainty principle [2,3], i.e. the spectral resolution of an eigenvalue is limited to $ \sim 1/ T$ for a wave packet evolved for a time interval $T$. However, the bound to the spectral resolution imposed by the time-energy uncertainty principle, which could be avoided using more elaborate methods [3,4], does not pose a relevant limitation for our purposes. The procedure used to extract the eigenvalue and the corresponding eigenvector of the BIC mode follows the method of Ref.[2], extended to include the presence of scattered (unbounded) states in the neighborhood of the BIC mode. Let us indicate by $E_\alpha$ ($\alpha=1,2,3,...$) and $E(q)$ the point and continuous spectrum of $\hat{H}$, with eigenvectors $|\phi_{\alpha} \rangle$ and $|\phi(q) \rangle$, respectively, forming an orthonormal basis. If the system is initially prepared in the state $| \psi(0) \rangle$, then the state vector evolves according to 
\begin{eqnarray}
|\psi(t) \rangle & = & \exp(-i \hat{H}t) | \psi(0) \rangle= \sum_{\alpha} a_{\alpha} |\phi_{\alpha} \rangle \exp(-i E_{\alpha} t) \nonumber \\
& + & \int dq \; a(q) \exp[-i E(q)t ] | \phi(q) \rangle 
\end{eqnarray}
\begin{figure}
\includegraphics[width=8cm]{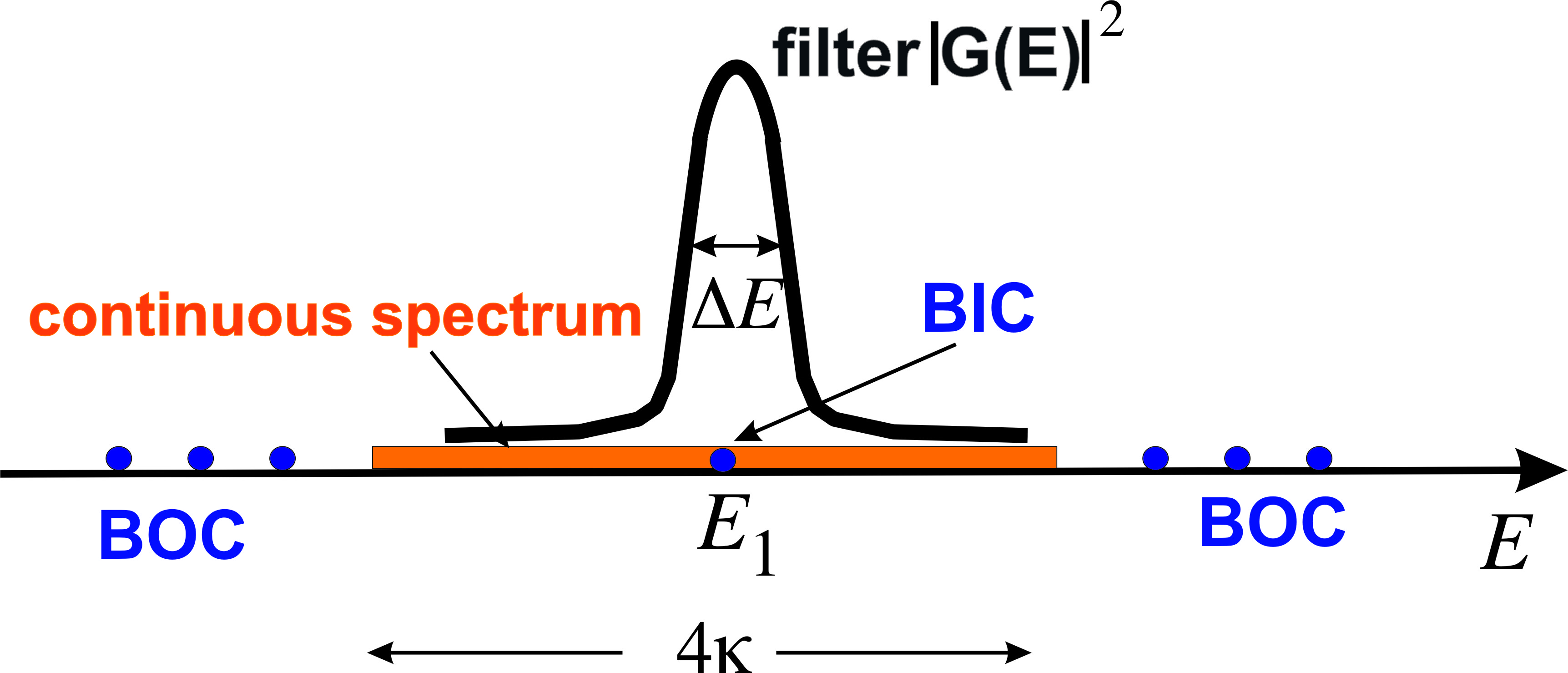}
\caption{(Color online) Schematic of the spectral filtering method to reconstruct the BIC mode with energy $E_1$ embedded in the continuous spectrum of scattered states.}
\end{figure}
\begin{figure}
\includegraphics[width=8.5cm]{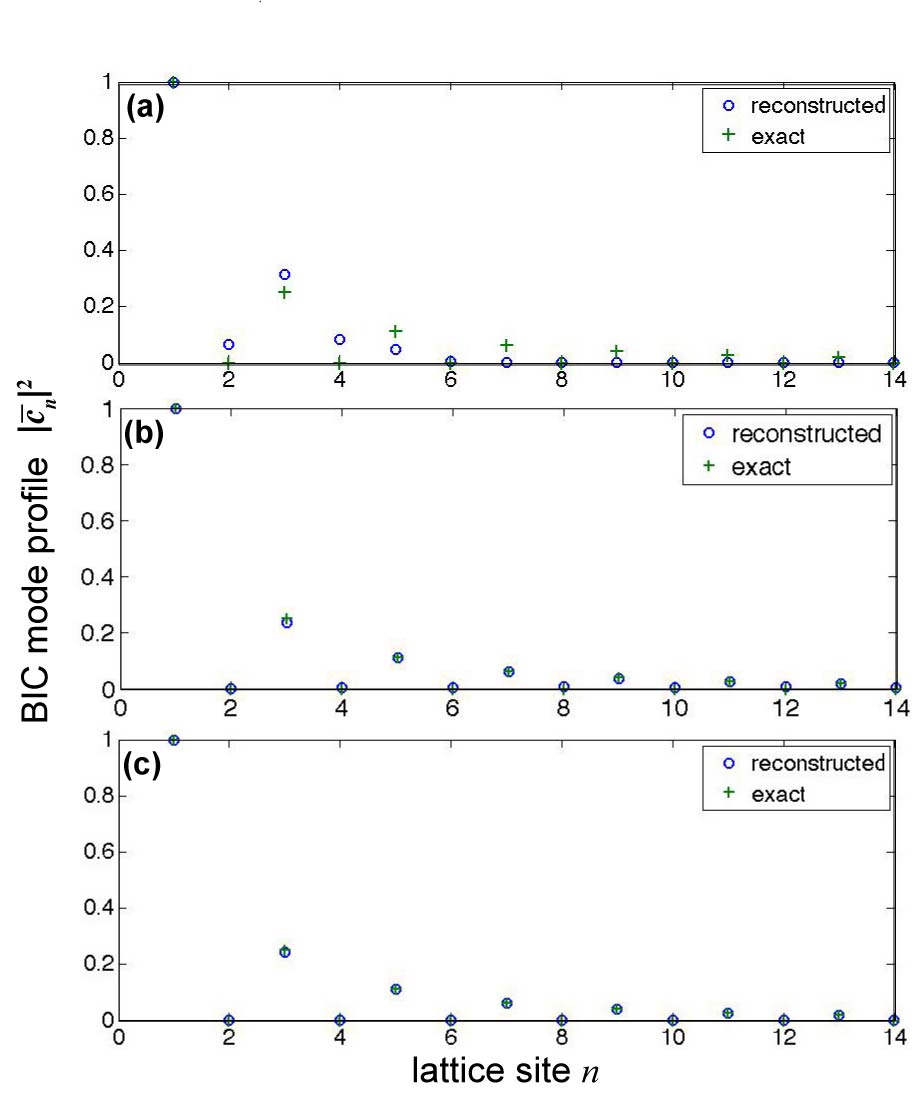}
\caption{(Color online) Reconstructed profile of the BIC mode at $E_1=0$, based on the inversion relation (S-17), for increasing values of $\kappa T$: (a) $\kappa T=2$, (b) $\kappa T=10$, and (c) $\kappa T=20$. The exact BIC mode profile is also shown in the figure for comparison.}
\end{figure}
where $a_{\alpha}= \langle \phi_{\alpha}| \psi(0) \rangle$, $a(q)= \langle \phi(q)| \psi(0) \rangle$ and $\sum_{\alpha} |a_{\alpha}|^2+\int dq |a(q)|^2=1$ for normalization.\\ 
To estimate the position of the discrete energy levels $E_{\alpha}$, let us consider the Fourier transform $\hat{C}(E)$ of the correlation function $C(t) =  \langle \psi(0) | \psi(t) \rangle$ with a temporal window function $g(t)$, i.e. [2]
\begin{equation}
\hat{C}(E)=\int_{0}^{T}dt \;  g(t) \langle \psi(0)| \psi(t) \rangle \exp(iEt)
\end{equation}
where $T$ is the maximum observation time. For the window function, we typically assume a square-wave function $g(t)=g_S(t)=1/T$ for $0<t<T$ and $g_S(t)=0$ otherwise, or the Gaussian function $g(t)=g_G(t)=(1/T) \exp[-(t-T/2)^2/w^2]$  for $0<t<T$ (with $w< \sim T/2$) and $g_G(t)=0$ otherwise. Substitution of Eq.(S-6) into Eq.(S-7) yields
\begin{equation}
\hat{C}(E)=\sum_{\alpha} | a_{\alpha}|^2 G(E-E_{\alpha})+\int dq |a(q)|^2 G(E-E(q))
\end{equation}
where $G(E) \equiv \int_{-\infty}^{\infty} dt \; g(t) \exp(iEt)$ is the spectral filter function.  For example, assuming for the sake of definiteness the square-wave window function $g(t)=g_S(t)$, one has 
\begin{equation}
G(E)= \exp(iET/2) \frac{\sin(ET/2)}{(ET/2)}.
\end{equation}
The spectral resolution of the method is determined by the banwidth of the filtering function $|G(E)|^2$, which is given by $\Delta E \simeq 2 \pi/T$. In fact, let us assume that we wish to estimate the positions of the eigenvalues $E_1$, $E_2$, $E_3$ ... of the bound states from the behavior of $|\hat{C}(E)|^2$. Let us first assume that there are not BIC modes, i.e. that $E_1$, $E_2$, $E_3$,... are not embedded into the continuous spectrum of scattered states. This case is discussed in Ref.[2]. If the observation time $T$ is long enough such that the resolution $\Delta E$ is smaller than the minimum separation of the discrete energy levels $E_{\alpha}$, from Eqs.(S-8) and (S-9) is follows that  $|\hat{C}(E)|^2$ shows a sequence of peaks at $E=E_{\alpha}$ of width $ \sim \Delta E$, provided that $|a_{\alpha}|$ is non-negligible, i.e. provided that the initial wave packet has a non-vanishing projection into the eigenvector $|\phi_{\alpha} \rangle$.  Hence the peaks of $|\hat{C}(E)|^2$ provide an estimation of the positions of the discrete energy levels with a resolution $\Delta E$ which is limited by the time-energy uncertainty principle, i.e. by the observation time $T$. The second case, which is of interest for our work, is the occurrence of a BIC mode with an energy $E_1$  embedded into the spectrum of scattered states, as sketched in Fig.5. In this case,  the first term on the right hand side of Eq.(S-8) shows a peak at $E \sim E_1$, however since the energy $E_1$ is buried into the continuous spectrum a contribution to $\hat{C}(E)$ arises as well as from the integral term on the right hand side of Eq.(S-8) owing to the finite bandwidth of the filtering function, as shown in Fig.5. The presence of the BIC mode can be thus properly estimated provided that the latter contribution to $\hat{C}(E)$ is smaller than the former one, i.e. provided that 
\begin{equation}
I \ll |a_1|^2
\end{equation}
 where we have set 
\begin{equation}
I = \left| \int dq |a(q)|^2 G(E_1-E(q) ) \right|.
\end{equation}
\begin{figure}
\includegraphics[width=8.3cm]{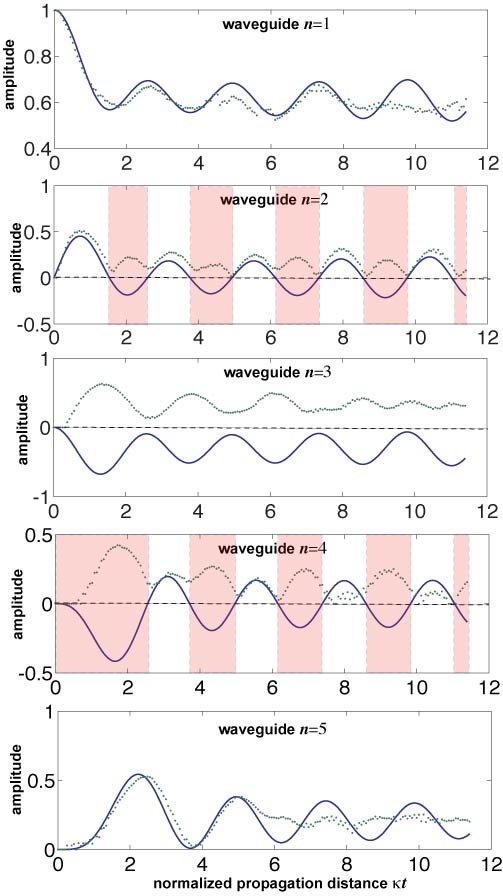}
\caption{(Color online) Evolution of the normalized amplitudes $|c_n(t)|$ along the the five waveguides $n=1,2,3,4$ and 5 near the lattice edge, as obtained from the experimental intensity map (dotted curves).  The solid thin curves show the behavior of $c_n(t)$ [${\rm Re}(c_n)$ for $n$ odd, ${\rm Im}(c_n)$ for $n$ even] as predicted by the coupled-mode equation model. The shaded areas show the time intervals where sign change is applied to the experimental curves in the reconstruction of the mode profile at even lattice sites.}
\end{figure}
Such a requirement is generally satisfied whenever the spectral resolution $\Delta E$ is much smaller than the bandwidth of the continuous spectrum. In fact, let us consider specifically the tight-binding lattice of Fig.1(a) given in the text, which applies to our experiment. In this case the continuous spectrum spans the energy range $(-2 \kappa, 2 \kappa)$, $q$ is the wave number of scattered states with the energy dispersion relation $E(q)=-2\kappa \cos q$ ($0 \leq q \leq \pi$), and there is one BIC mode at energy $E_1=0$, i.e. at the center of the band of scattered states. An upper limit to $I$ can be obtained from the Cauchy-Schwarz inequality
\begin{eqnarray}
I^2  & \leq & \int dq |a(q)|^2 \int dq |G(E-E(q))|^2 \nonumber \\
& < & \int dq |G(E-E(q))|^2
\end{eqnarray}
For an observation time $T$ long enough such that the spectral resolution $\Delta E$ is much smaller than the bandwidth $4 \kappa$ of the continuous spectrum,  i.e. for $\kappa T \gg 1$, one can assume $|G(E)|^2 \simeq ( 2 \pi /T) \delta (E)$, so that one obtains
\begin{equation}
I^2 < \frac{2 \pi }{T}\int dq \delta(E-E(q))= \frac{2 \pi}{T |(dE/dq)_{q_1}|}
\end{equation}
with $E(q_1)=E_1$. For $E_1=0$ one has $|(dE_1/dq)|_{q_1}=2 \kappa$ and thus $I < \sqrt{\pi /(T \kappa)}$.  
Note that, since $\kappa T \gg 1$, one has $I \ll 1$ and thus the inequality (S-10) is readily satisfied, provided that the initial wave packet has a projection $|a_1|^2$ onto the BIC mode of order $\sim 1$. Note also that the contribution to $\hat{C}(E)$ around $E=E_1$ arising from the BOC modes with energies $E_2$, $E_3$, ... outside the continuos band is negligible for $E \sim E_1$ since the spectral filter $G(E)$ has a width $\Delta E$ much smaller than $4 \kappa$ (see Fig.5).\\
Once the position of the energy level $E_{\alpha}$ has been estimated from the peaks of $|\hat{C}(E)|^2$, the corresponding eigenvector $| \phi_{\alpha} \rangle$ can be approximately reconstructed by considering the Fourier transform of the evolved wave packet with the time window $g(t)$, i.e. 
\begin{equation}
| \psi (E) \rangle = \int_{- \infty}^{\infty}dt g(t) \exp(-iEt) | \psi(t). \rangle
\end{equation} 
Using Eq.(S-6), Eq.(S-14) can be cast in the form
\begin{equation}
| \psi (E) \rangle = \sum_{\alpha} a_{\alpha} G(E-E_{\alpha}) | \phi_{\alpha} \rangle + \int dq a(q) G(E-E(q)) | \phi(q) \rangle.
\end{equation} 
For $E=E_1$ (the energy level of the BIC mode) and assuming again $ \kappa T \gg 1$, the contribution in the sum on the right hand side of Eq.(S-15) comes from the BIC mode solely, whereas the contribution to $| \psi_E \rangle$ arising from the BOC modes and from the scattered states can be again neglected (provided that $|a_1|^2 \sim 1$). This can be proven following the same procedure given above, calculating the norm of the continuous spectrum component to the vector $|\psi(E) \rangle$  and using the Cauchy-Schwarz inequality. 
Hence the BIC eigenvector $|\phi_1 \rangle$ with energy $E=E_1$ can be reconstructed using the relation
\begin{equation}
a_1 | \phi_{1} \rangle \simeq | \psi(E_1) \rangle=\int_{- \infty}^{\infty} dt g(t)  \exp(iE_1t) | \psi(t) \rangle.
\end{equation}
{\it {\bf S.4. BIC mode reconstruction.}}\\
In the Wannier basis representation, apart from a normalization factor Eq.(S-16) enables to reconstruct the occupation amplitudes $\bar{c}_n(E_1)$ of the BIC mode from the knowledge of the evolved wave packet amplitudes $c_n(t)$ by means of the inversion relation
\begin{equation}
\bar{c}_n (E_1)  \simeq \int_{- \infty}^{\infty} dt g(t) \exp(-iE_1t) c_n(t).
\end{equation}
Excitation of the boundary lattice site, i.e. assuming the initial condition $c_{n}(0)=\delta_{n,1}$,  ensures that the wave packet projection  onto the BIC mode is of order $\sim 1$  [see Fig.2(b) in the main text], wheres BOC states are weakly excited. To estimate the propagation time $T$ needed to accurately reconstruct the BIC eigenvector using the inversion relation (S-17), in Fig.6 we show the reconstructed eigenvector $|\bar{c}_n(E_1)|^2$ for a few increasing values of $\kappa T$, obtained from Eq.(S-17) with $E_1=0$ and with amplitudes $c_n(t)$ numerically computed by solving the coupled-mode equations (S-3) with the initial condition $c_n(0)=\delta_{n,1}$. The reconstructed eigenvector is compared with the predicted one, i.e. $\bar{c}_n=0$ for $n$ even and $|\bar{c}_n| \propto  2/(n+1)$ for $n$ odd. Note that, as expected, the accuracy in the reconstructed eigenvector increases as $\kappa T$ is increased. Note also that for a propagation time $T= 10 / \kappa$ the inversion relation (S-17)  reproduces the BIC mode with great accuracy. This case corresponds basically to our experimental condition, where beam propagation along the full sample length corresponds indeed to an observation time interval $T \simeq 11.4 / \kappa$.\\
In our experiment, the fluorescence imaging method [5,6] was used  to accurately measure the evolution of the light field intensity distributions in the various waveguides. Femtosecond laser writing in fused silica creates color centers that provide fluorescent emission at about 650 nm, when light at 633 nm is propagated in the waveguide. Top-view imaging of the fluorescence signal is employed to visualize and quantitatively estimate the light distribution along the waveguide array, rejecting the background
light by a notch filter at 633 nm. Light collection for the top-view images is performed with a $\times10$ objective, 0.25 numerical aperture, placing the plane of arrayed waveguides
 at focus. To achieve a high resolution in
imaging the light in the waveguide array all along its length, several images have
been acquired and then stitched together. Propagation losses in the waveguides
have been compensated by re-normalizing the intensity levels in the acquired
images. The resulting map, showing the propagation of the beam intensity along the array, is shown in Fig.2(c) of the main text. From the measured map, the evolution of $|c_n(t)|$, i.e. of the square root of normalized light intensity trapped in waveguide $n$, can be then retrieved. Figure 7 shows the detailed behavior of $|c_n(t)|$ as obtained from the measured intensity map for the five waveguides near the array edge. It should be noted that the application of the spectral inversion relation (S-17) requires the knowledge of both modulus and phase of $c_n(t)$. Fortunately, for our system the measure of the phase of $c_n(t)$ was not necessary. In fact, from Prop. IV  stated in S.1 it turns out that, for the given initial excitation, $c_{n}(t)$ is real-valued for $n$ odd and purely imaginary for $n$ even. Therefore, apart from an unimportant phase at alternating sites, one can  set $c_{n}(t)= \pm |c_n(t)|$ on the right hand side in Eq.(S-17). In Fig.7 we show the theoretical behaviors of $c_n(t)$ for the five waveguides near the array edge. To reconstruct the profiles $c_n(t)$ from the measured curves $|c_n(t)|$, special attention should be paid to possible sign alternations of $c_{n}(t)$, which occur when $c_n(t)$ oscillates crossing zero. In this case sign alternation should be properly included into the measured curves $|c_n(t)|$. In particular, for odd sites $n$, it turns out that one can safely set $c_n(t)=|c_n(t)|$  for $n < \sim 13$, i.e. sign alternation is not necessary because $c_n(t)$ does not change sign (see, for instance, the theoretical curves in Fig.7 for $n$ up to 5). Conversely, sign alternations should be introduced at even sites because of the oscillatory behavior of $c_n(t)$  (see Fig.7 for waveguides $n=2$ and $n=4$). Using this procedure, we could thus reconstruct the profile $|\bar{c}_n(E_1)|^2$ of the BIC mode by means of the spectral inversion relation (S-17). The reconstructed BIC profile is shown in Fig.3(b) given in the main text.\\
\\  
\noindent
{[1]} M.I. Molina, A.E. Miroshnichenko, and Y.S. Kivshar,
Phys. Rev. Lett. {\bf 108}, 070401 (2012).\\
{[2]} M. D. Feit, J. A. Fleck, and A. Steiger, J. Comp. Phys. {\bf 47}, 412 (1982).\\
{[3]} D. Neuhauser, J. Chem. Phys. {\bf 93}, 2611 (1990).\\
{[4]} M. R. Wall and D. Neuhauser, J. Chem. Phys. {\bf 102}, 8011 (1995).\\
{[5}] A. Szameit, F. Dreisow, H. Hartung, S. Nolte, A. T\"{u}nnermann, and F. Lederer, Appl. Phys. Lett. \textbf{90}, 241113 (2007).\\
{[6]} G. Corrielli, A. Crespi, G. Della Valle, S. Longhi, and R. Osellame, Nature Comm. {\bf 4}, 1555 (2013).

\end{document}